\documentclass[%
superscriptaddress,
 preprint,
showpacs,preprintnumbers,
amsmath,amssymb,
aps,
prb,
]{revtex4-2}

\usepackage{graphicx}
\usepackage{dcolumn}
\usepackage{bm}
\usepackage{amsmath}
\usepackage{upgreek}
\usepackage{hyperref}

\begin{document}


\title{Edge-enhanced diffractive neural networks based on spin-multiplexed nonlocal metasurfaces}

\author{Qianqian He}
\affiliation{School of Information Engineering, Nanchang University, Nanchang 330031, China}

\author{Kenan Guo}
\affiliation{School of Information Engineering, Nanchang University, Nanchang 330031, China}
\affiliation{National Laboratory of Solid-State Microstructures and College of Engineering and Applied Sciences, Nanjing University, Nanjing 210093, China}

\author{Jumin Qiu}
\affiliation{School of Information Engineering, Nanchang University, Nanchang 330031, China}

\author{\mbox{Huifu Qiu}}
\affiliation{National Laboratory of Solid-State Microstructures and College of Engineering and Applied Sciences, Nanjing University, Nanjing 210093, China}

\author{Shuyuan Xiao}
\email{syxiao@ncu.edu.cn}
\affiliation{School of Information Engineering, Nanchang University, Nanchang 330031, China}
\affiliation{Institute for Advanced Study, Nanchang University, Nanchang 330031, China}

\author{Tingting Liu}
\email{ttliu@ncu.edu.cn}
\affiliation{School of Information Engineering, Nanchang University, Nanchang 330031, China}
\affiliation{Institute for Advanced Study, Nanchang University, Nanchang 330031, China}

\begin{abstract}
Single-layer diffractive neural networks often face classification accuracy bottlenecks due to limited wavefront modulation capabilities. Edge detection, as an optical image processing technique, extracts image contours and offers a promising way to simplify classification tasks. However, integrating edge detection and DNN-based classification on a single chip remains a challenge. Here, we propose an integrated nonlocal meta-platform that sequentially achieves all-optical edge detection and classification via spin-multiplexing. By exploiting the dispersion properties of the nonlocal Huygens' metasurface, the co-polarized component in the output light performs momentum-space filtering for real-time edge detection. Subsequently, the cross-polarized component undergoes geometric phase modulation to execute image classification within the DNN. We couple quasi-bound states in the continuum and magnetic dipole resonances in crescent-shaped nanopillars, achieving a high polarization conversion efficiency of approximately 55\%. This edge-enhanced DNN architecture significantly reduces data redundancy, elevating the classification accuracy of the single-layer network on the MNIST dataset from 64.2\% to 80.7\%. Our work provides a compact, high-efficiency solution for integrated all-optical machine vision and intelligent photonic computing. 
\end{abstract}

\maketitle


\section{Introduction}

Diffractive neural networks (DNNs) have emerged as a transformative paradigm for optical computing. These networks are physically formed by diffractive surfaces that work in collaboration to optically perform an arbitrary function that the network can statistically learn, offering unparalleled advantages in processing speed and energy efficiency\cite{hu2024diffractive,li2025optimized,sun2023review,omar2025from}. Since the architecture of DNNs was proposed, extensive efforts have been dedicated to enhance their classification accuracy and functional flexibility, through strategies such as residual connections, differential detection and multiplexing in wavelength, polarization, and orbital angular momentum\cite{dou2020residual,lu2024metasurface,li2019class,li2021spectrally,duan2023optical,luo2022metasurface,qiu2026integrated,li2022polarization,cheng2026orbital,wang2022diffractive,he2026orbital,liu2025dual}. However, a fundamental trade-off persists between system compactness and computational performance. Multilayer architectures can theoretically enhance accuracy, but they introduce stringent inter-layer alignment requirements and inevitable error accumulation. Conversely, single-layer DNNs, despite their extreme structural simplicity, frequently encounter an accuracy bottleneck when processing large-scale datasets due to the absence of network depth and nonlinear activations\cite{yu2025nonlinear,raghu2017}.

To address this limitation, integrating task-specific preprocessing steps, particularly edge detection, presents a promising pathway to simplify subsequent classification tasks\cite{nikolay2026analog,cui2026metasurfaces}. Previous studies have validated the effectiveness of such preprocessing operations. For instance, learnable structured illumination or a specific wavelength pair was introduced to perform complementary encoding of the input signal, combined with a differential detection mechanism to expand the network's decision domain\cite{zhang2024advanced,wang2026all}. While this improves performance, it significantly increases system complexity. Another approach is to construct an optical diffractive convolutional neural network\cite{abdoll2020metaoptics,yu2023optical,liang2026metabased}, where a convolutional layer performs effective feature extraction before classification, improving the classification accuracy. But this comes with demanding fabrication and alignment requirements. Notably, by extracting object contours and significantly compressing redundant spatial data, edge-enhanced imaging can alleviate the computational load on the diffractive layers. Nevertheless, conventional optical edge detection remains reliant on a 4f system, which is bulky and requires complex alignment\cite{yang2021switchable,tu2023optical}. Alternatively, designing metasurface phase profiles as spatial differential operators\cite{yin2026terahertz,yang2023realizing,zhou2019optical,tu2025inverse}, also faces challenges related to multi-device alignment and optical integration. All these schemes necessitate the fabrication of separate devices, which present formidable challenges for monolithic integration with DNN frameworks.

Nonlocal metasurfaces provide a novel approach to addressing the aforementioned challenges\cite{ding2018gradient,he2018high,he2020meta,ullah2022recent,zou2024advanced}. By engineering periodic arrays to excite collective resonances among the unit cells, such devices can exhibit momentum-space filtering characteristics, which have been successfully exploited for real-time edge detection\cite{kwon2018nonlocal,komar2021edge,jiang2025research}. In parallel, by spatially manipulating the resonant elements to engineer the Pancharatnam-Berry (PB) phase, metasurfaces can provide arbitrary wavefront modulation\cite{malek2022multifunctional,overvig2021chiral,liu2024phase}. This mechanism enables edge detection and holography within a single device\cite{yu2026double}. In recent years, the deliberate breaking of structural symmetry has introduced the designs based on quasi-bound states in the continuum (q-BIC)\cite{zhou2025dual,liu2024edge,zhang2025momentum,zhou2025ultrahigh,koshelev2018asymmetric,huang2023leaky,li2026multimodal}, enabling efficient functionalities such as metalens imaging\cite{yao2025nonlocal,ouyang2025ultra}. While efficient wavefront shaping and edge detection have been demonstrated separately, the monolithic integration of these functionalities remains rarely explored. Achieving all-optical edge detection preprocessing followed by DNN-based classification on a single, compact metasurface platform poses substantial challenges. Such integration requires not only the spatial decoupling of the edge-enhanced optical signal from the classification channel but also a high polarization conversion efficiency to ensure high signal-to-noise ratios, which conventional designs struggle to satisfy simultaneously.	

In this work, we propose a nonlocal Huygens' metasurface architecture that achieves all-optical integration of edge detection and DNN-based classification via time-sequential spin-multiplexing. In our scheme, the co-polarized component in the output light exploits the dispersion properties of the nonlocal metasurface to perform edge detection. Subsequently, the cross-polarized component undergoes geometric phase modulation to perform image classification. We employ a crescent-shaped silicon nanopillar design that couples q-BIC with magnetic dipole resonances (MDR), resulting in a high polarization conversion efficiency of approximately 55\%, surpassing the efficiency typically limited to approximately 25\%. Our results show that this integrated edge detection preprocessing significantly enhances the classification accuracy for the MNIST dataset from 64.2\% to 80.7\%. This architecture provides a promising solution for real-time visual perception, biomedical morphological analysis, and high-precision industrial inspection.

\section{Results and Discussion}
\subsection{Principle and metasurface design}

Fig. 1 illustrates a single-layer nonlocal Huygens' metasurface. With an input digit image under right circularly polarized (RCP) light illumination, the unconverted (co-polarized) RCP light exploits the dispersion properties of the nonlocal metasurface for momentum-space filtering, which extracts edge features from the input image to generate an edge-enhanced result, thereby completing the optical preprocessing. Reusing this generated edge-enhanced image as the new input, the converted (cross-polarized) left circularly polarized (LCP) light is modulated by the metasurface's geometric phase profile to perform image classification within the DNN. We integrate optical edge detection and DNN-based classification on a single metasurface through time-sequential spin-multiplexing. Compared to a cascaded two-layer configuration, our scheme offers significant advantages in terms of alignment difficulty and system cost. This design utilizes the metasurface's nonlocal properties to achieve high-speed, low-power edge detection preprocessing on images. Using this preprocessed output as the input to the subsequent DNN for classification, this edge-enhanced DNN architecture significantly improves the classification performance of the single-layer network when handling large-scale datasets. 

\begin{figure*}[htbp]
	\centering
	\includegraphics[width=0.9\textwidth]{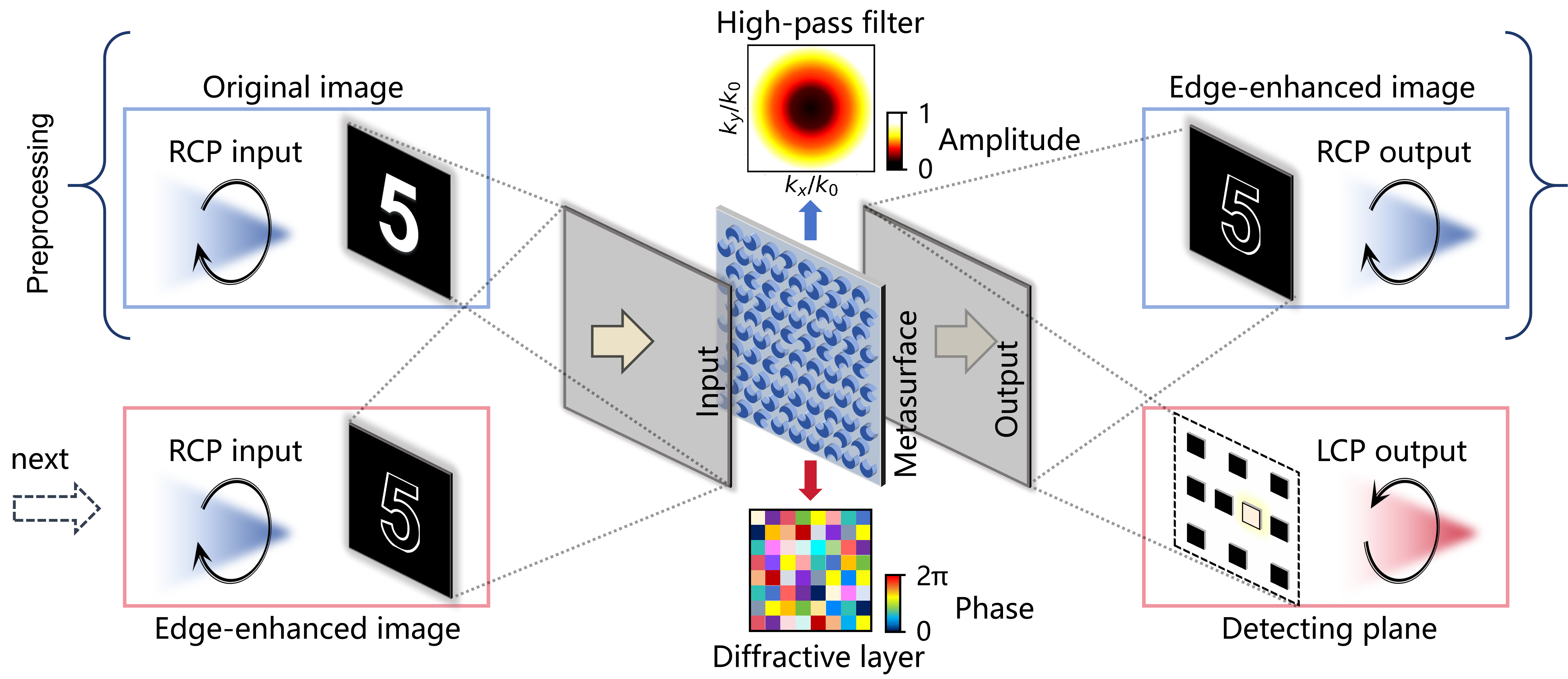}
	\caption{Schematic of the edge-enhanced DNN architecture, where a nonlocal Huygens’ metasurface enables edge-enhanced imaging (preprocessing task) and DNN-based classification (main task) via time-sequential spin-multiplexing.}
	\label{fig1}	
\end{figure*}

To achieve the dual functionalities illustrated in Fig. 1, we carefully design the metasurface unit cell. The specific configuration and structural parameters are shown in Fig. 2(a). Crescent-shaped silicon nanopillars are arranged in a hexagonal lattice on a glass substrate. Each nanopillar is formed by etching a smaller cylinder of diameter $D_2$, offset by a distance $L$, from a larger cylindrical pillar with diameter $D_1$. The array period is fixed at $P = 1000$ nm, the unit height at $H = 327$ nm, and the large cylinder diameter at $D_1 = 620$ nm. The offset $L$ controls the degree of asymmetry, while the small cylinder diameter $D_2$ primarily affects the spectral position and linewidth of MDR. We optimize these two parameters using the finite difference time domain (FDTD) method, aiming to simultaneously achieve: (1) high polarization conversion efficiency $T_{\text{RL}}$ (the transmittance ratio $T_{\text{RL}}$ ($T_{\text{RR}}$) is defined as the output optical power of LCP (RCP) light to the incident optical power of RCP); (2) phase modulation capability covering the full 2$\pi$ range; and (3) strong angular dispersion induced by collective resonances. The final optimized parameters are $L = 220$ nm and $D_2 = 470$ nm.

\begin{figure*}[htbp]
	\centering
	\includegraphics[width=0.9\textwidth]{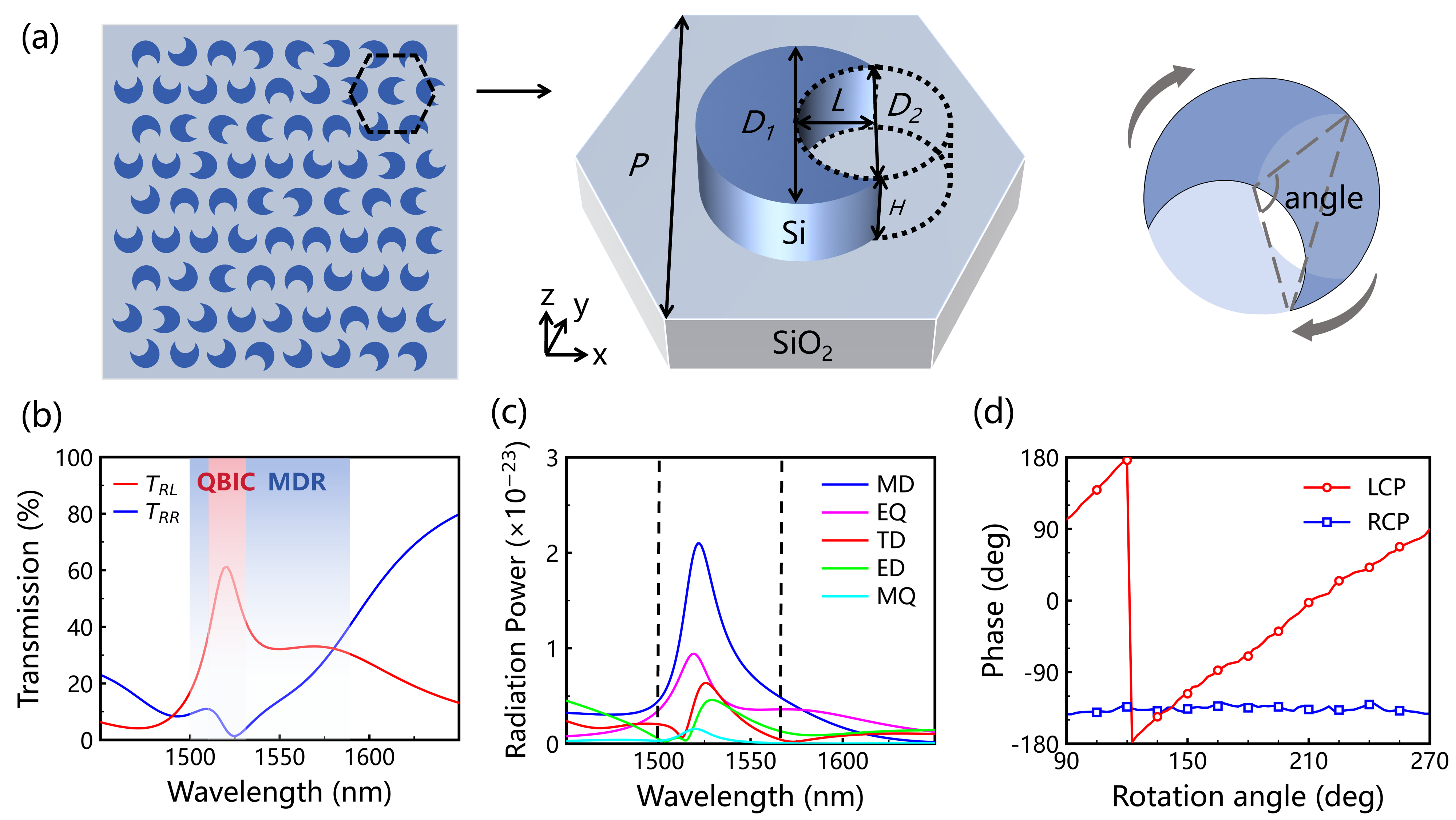}
	\caption{Unit-cell design and optical response of the metasurface. (a) Crescent-shaped silicon nanopillars arranged in a hexagonal lattice on a glass substrate. The geometric phase is encoded by rotating the nanopillars. (b) Transmission spectra of the converted LCP and the unconverted RCP light. The red and blue regions indicate the bandwidths of q-BIC and MDR. (c) Multipole decomposition of the designed resonant unit cell. (d) Phase distributions of the two output light components as a function of the nanopillar rotation angle.}
	\label{fig2}
\end{figure*}

Under the optimized set of parameters, we further simulate the transmission characteristics of the resonant unit cell. The transmission spectra of the LCP and RCP light in the wavelength range of 1450--1650 nm are shown in Fig. 2(b). It can be observed that the cross-polarized transmission $T_{\text{RL}}$ exhibits a sharp, narrow peak at 1524 nm, while the co-polarized transmission $T_{\text{RR}}$ is strongly suppressed to a near-zero dip at the same wavelength. This arises from a q-BIC resonance (red region) excited by the structural asymmetry, which is further validated by a topological charge analysis presented in Sec. 4 of the Supplemental Material\cite{SM}. Additionally, $T_{\text{RL}}$ shows another broad and flat peak at 1540 nm. According to the multipole decomposition shown in Fig. 2(c), this peak originates from MDR (blue region). The optimized parameters cause the resonance peaks of q-BIC and MDR to overlap spectrally. Their interference generates a Fano lineshape, which satisfies the generalized Kerker condition for efficient forward scattering. $T_{\text{RL}}$ reaches a peak of approximately 55\%. This high efficiency ensures sufficient optical power is converted into the LCP channel. Due to rotational asymmetry, as shown in Fig. 2(d), the phase of the output LCP light varies linearly over the full 2$\pi$ range as the nanopillar rotation angle changes from $90^\circ$ to $270^\circ$, allowing the realization of arbitrary phase profiles for the metasurface based on the PB phase principle. The enhanced efficiency and stable phase modulation for output LCP light contribute to efficient DNN-based image classification. In contrast, the phase of the output RCP light remains nearly flat, introducing no phase perturbation related to the spatial rotation angle. Thus, its flat wavefront does not interfere with the complex wavefront encoded by the geometric phase in the LCP channel. The two output polarization states are orthogonal, and the two functionalities rely on two independent dimensions: amplitude and phase. Consequently, the two channels can independently perform their respective functions. Considering potential fabrication errors in practical validation, we further analysis the influence of structural parameter perturbations on edge detection and classification performance. For details, see Sec. 5 of the Supplemental Material\cite{SM}.

\subsection{Optical edge-enhanced imaging}

We first characterize the edge detection performance of the metasurface. As analyzed previously, our design ensures minimal phase change for RCP light under geometric rotation, meaning the functionality does not rely on phase modulation. We analyze its amplitude response, specifically the influence of the incident angle on the transmission spectra of the co-polarized light. As shown in Fig. 3(a), at normal incidence, $T_{\text{RR}}$ exhibits a near-zero dip at the resonance wavelength of 1524 nm. Under oblique incidence, the resonance wavelength undergoes a significant red shift as the incident angle increases. Fig. 3(b) further reveals that at the resonance wavelength, $T_{\text{RR}}$ grows quadratically as the incident angle increases from $0^\circ$ to $16^\circ$, fitting well with an ideal second-order differentiator response (red dashed line). This trend originates from the resonance redshift phenomenon shown in Fig. 3(a), which demonstrates the existence of strong angular dispersion, a hallmark feature of nonlocal resonance. Consequently, at the original operating wavelength, $T_{\text{RR}}$ increases, reaching up to approximately 60\% under oblique incidence. This angular response defines an effective numerical aperture (NA) of about 0.27, corresponding to an edge resolution of 3.4 $\upmu\mathrm{m}$, which quantifies the ability of the metasurface to resolve fine details in imaging.

\begin{figure*}[htbp]
	\centering
	\includegraphics[width=0.9\textwidth]{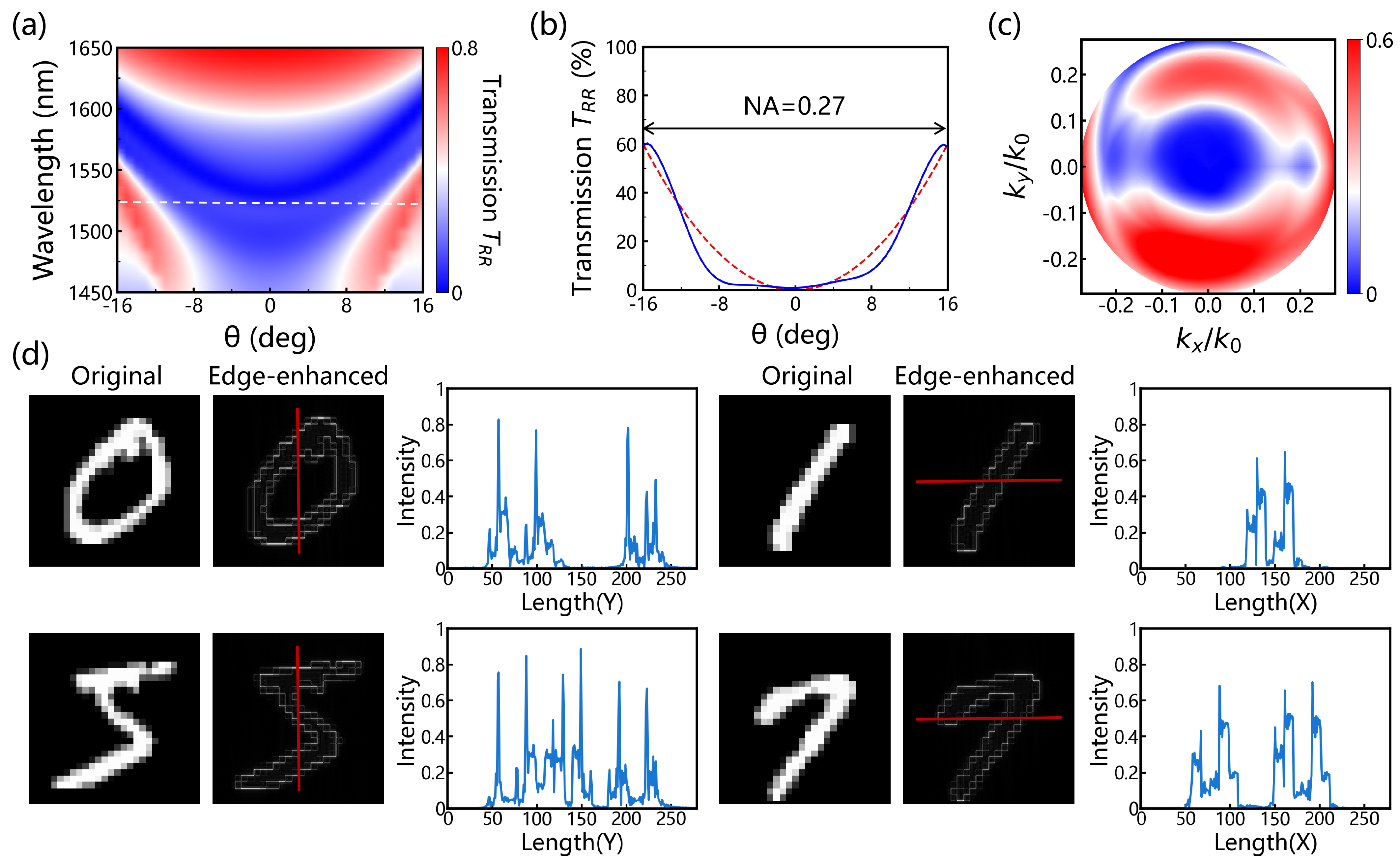}
	\caption{Optical responses of the output unconverted RCP light and edge detection performance of the metasurface. (a) $T_{\text{RR}}$ as a function of both wavelength and incident angle. (b) $T_{\text{RR}}$ variation with incident angle at the resonant wavelength, along with its quadratic fit. (c) 2D transmission dispersion map at the operating wavelength. (d) Comparison between original and edge-enhanced images of several digits (0, 5, 1, 7) from the MNIST dataset, and the corresponding intensity profiles along the red solid lines marked in the edge-enhanced images.}
	\label{fig3}
\end{figure*}

To obtain the optical transfer function for edge detection, we extend the analysis from one-dimensional angular scanning to the two-dimensional (2D) momentum space. Fig. 3(c) displays the 2D transmission dispersion map of $T_{\text{RR}}$ as a function of both incident angle and azimuthal angle. The map clearly shows that the transmission in the central region is strongly suppressed to near-zero, while that in the peripheral region remains at a relatively high level. This characteristic enables the metasurface to transmit the high spatial frequencies that carry object edge information, while blocking the low spatial frequencies associated with the background. Moreover, as shown in Fig. 2(d), the phase response for the RCP light remains flat. This avoids stray diffraction caused by a non-uniform phase profile, which would otherwise lead to image blurring. The realization of the edge detection function relies entirely on the high-pass filtering characteristic presented by the transmission of the unconverted RCP light. Combining these response characteristics, the output RCP light is suitable for edge-enhanced imaging.

Next, we show the edge-enhanced imaging results of the designed metasurface. Fig. 3(d) compares the original images and edge-enhanced images of several digits (0, 5, 1, 7) from the MNIST dataset. The results show that the edges of the digits are clearly extracted after processing by the metasurface. To evaluate the edge detection quality, we further provide the vertical or horizontal intensity profiles of the output images. The curves exhibit sharp intensity peaks at the expected edge locations, with the background strongly suppressed and a high contrast between edges and background. This demonstrates that the metasurface achieves high-quality 2D edge detection.

\subsection{Metasurface-based DNN classification}

Fig. 4(a) illustrates the single-layer DNN architecture, which consists of an input plane, a single diffractive layer, and an output plane. The edge-enhanced image output from the RCP channel is encoded as the phase information of the input optical field. According to the Rayleigh–Sommerfeld diffraction theory, the light propagates forward in free space. After being modulated by the neurons of the diffractive layer, the loss is calculated at the output plane. The error gradients are then backpropagated along the optical path via the chain rule to iteratively update the phase of each neuron in the diffractive layer, eventually yielding a phase profile optimized for edge-enhanced images. The light propagation between layers is accurately modeled by the angular spectrum method; details are provided in the Sec. 1 of the Supplemental Material\cite{SM}. In the output LCP channel, the geometric rotation of the resonant elements provides stable phase modulation covering the full 2$\pi$ range for the LCP light. We design the phase response of the metasurface to match the optimized diffraction-layer phase distribution of the single-layer DNN, as shown in Fig. 4(b). This enables the metasurface to serve as the physical diffraction layer, processing the input optical field to achieve image classification. The amplitude response corresponds to the polarization conversion efficiency. From Fig. 2(b), $T_{\text{RL}}$ is approximately 55\%, providing sufficient optical power for diffractive propagation. A detailed analysis of the optical transmission efficiency of the entire architecture and its impact on classification performance is presented in Sec. 2 of the Supplemental Material\cite{SM}. Therefore, we design a complete single-layer edge-enhanced DNN architecture on a multifunctional metasurface device.

\begin{figure}[htbp]
	\centering
	\includegraphics[width=0.6\textwidth]{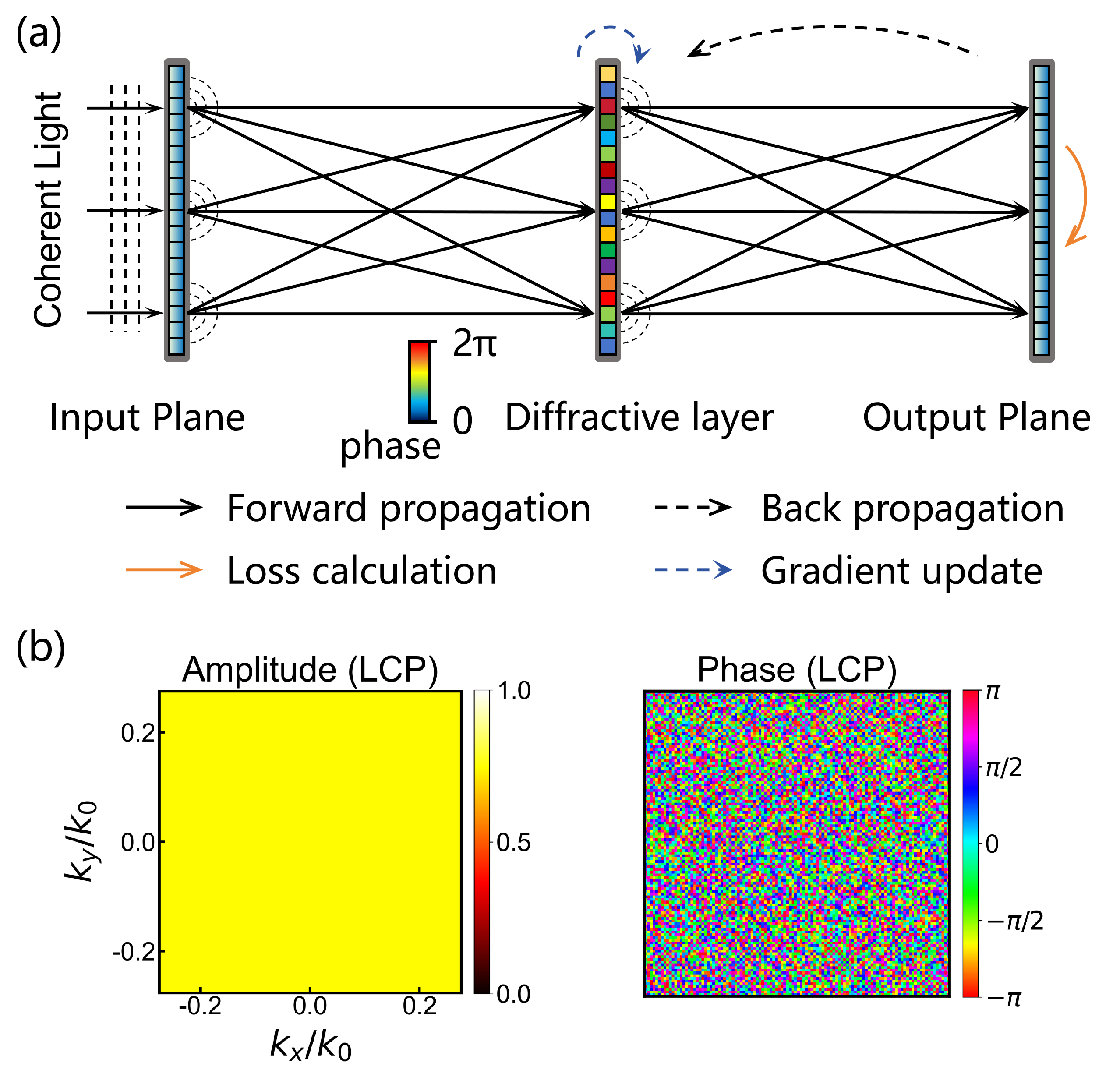}
	\caption{Schematic of the single-layer edge-enhanced DNN for image classification. (a) Architecture of the single-layer DNN, where the metasurface serves as the diffractive layer, and the input optical field carries the edge information of the image. (b) Amplitude and phase profiles of the designed metasurface corresponding to the converted LCP for image classification.}
	\label{fig4}
\end{figure}

We finally characterize the image classification performance of the metasurface. Through comparison, we verify the effectiveness of edge detection preprocessing in improving the classification performance of the single-layer DNN, while also considering the influence of the diffractive layer size (i.e., the number of neurons). The results are shown in Fig. 5(a). As the number of neurons in the diffractive layer increases, the classification accuracy improves modestly, but the increase does not scale linearly with size, because the performance gain from increased model capacity saturates. Notably, across all sizes, the classification accuracy with edge detection preprocessing is significantly higher than that with original inputs, strongly demonstrating the necessity and effectiveness of the preprocessing.

\begin{figure*}[htbp]
	\centering
	\includegraphics[width=0.9\textwidth]{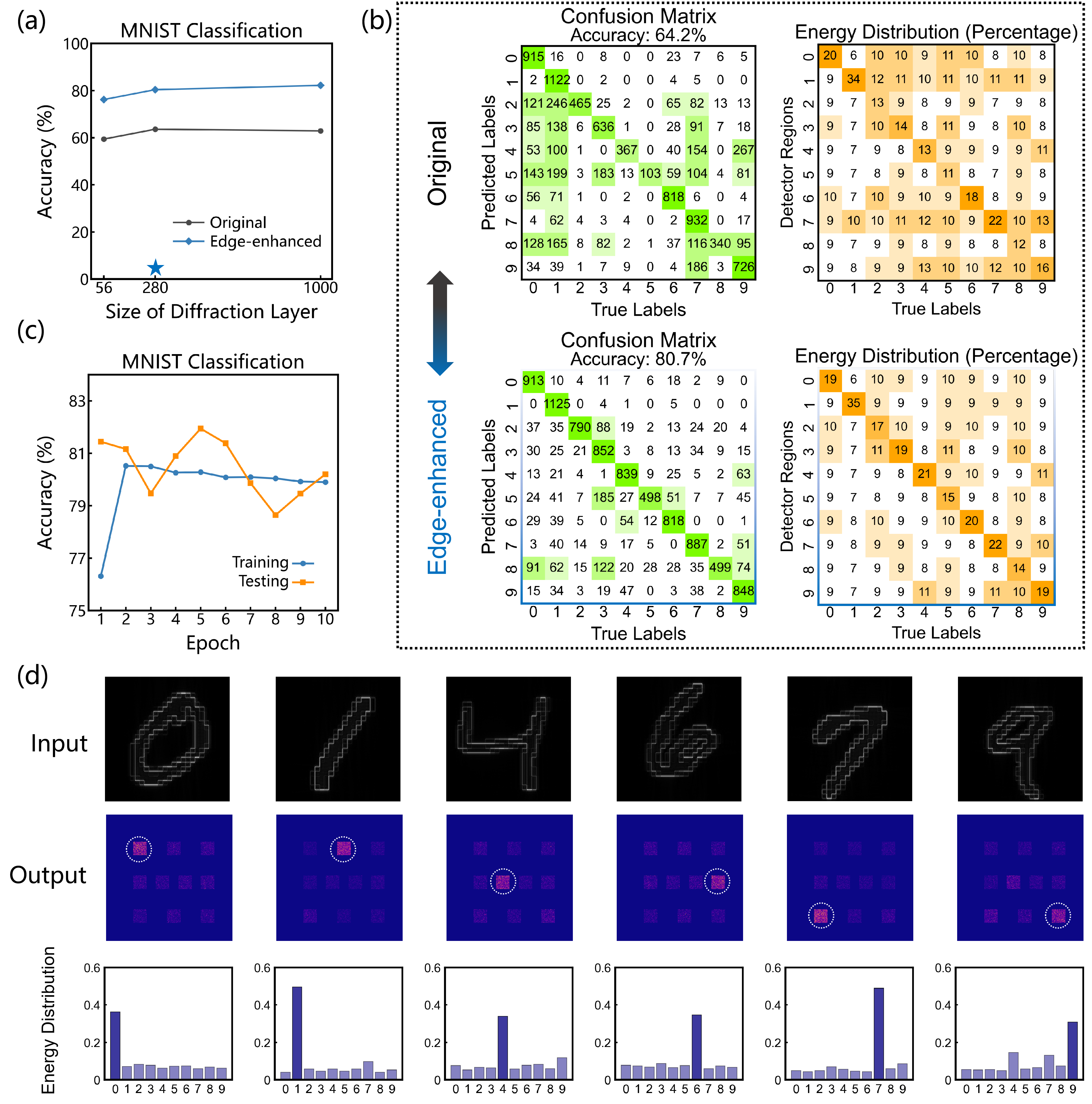}
	\caption{Image classification performance of the DNN. (a) Comparison of classification accuracy with and without edge detection preprocessing for MNIST, for diffractive layer sizes of $56\times 56$, $280\times 280$, and $1000\times 1000$. (b) Confusion matrices and energy distribution percentage matrices for original and edge-enhanced inputs at a diffractive layer size of $280\times 280$. (c) Training and testing accuracies for edge-enhanced input. (d) Several input digit images (top) and corresponding output results (middle) for the edge-enhanced DNN, where the output plane contains ten detector regions each corresponding to a digit class, with the final result marked by a white dashed circle, and the normalized energy distribution maps (bottom).} 
	\label{fig5}
\end{figure*}

To balance system compactness and classification accuracy, we present the final performance of the DNN with a size of $280\times 280$. In this configuration, the size of each detector area is $30\times 30$ pixels. As shown in Fig. 5(b), the network achieves a classification accuracy of 64.2\% on original inputs, reflecting its basic ability to learn classification. However, the optical energy distribution is relatively scattered, indicating noticeable classification confusion. When edge detection is introduced as preprocessing, the diagonality of the confusion matrix and the energy distribution matrix is significantly enhanced. The classification accuracy increases to 80.7\%, and the network concentrates optical energy more precisely onto the target detectors, greatly reducing classification confusion. This improvement benefits from edge detection, which extracts the contours of the digits and reduces data redundancy. This process significantly simplifies the classification task of the single-layer network while greatly improving computational efficiency, thereby ultimately leading to an effective increase in classification accuracy. Furthermore, Sec. 3 of the Supplemental Material provides a detailed analysis of how detector area and arrangement pattern affect accuracy\cite{SM}. To validate the robustness of the architecture in practical environments, we further compare the accuracy under different SNRs, detailed in Sec. 6 of the Supplemental Material\cite{SM}. Fig. 5(c) shows that the network performs consistently on the training and testing sets, indicating good fitting.

We perform simulation tests on this edge-enhanced DNN. On the output plane, the digit category corresponding to the region with the highest optical energy (marked by a white dashed circle) is taken as the classification result of the network, as shown in the middle row of Fig. 5(d). The energy distributions of the ten target regions are shown in the bottom row for better visualization. It can be observed that the network accurately focuses the optical energy of the input edge-enhanced digit onto the corresponding detector, with the optical energy in the target region significantly higher than that in other regions. To further verify the generality of the architecture, our test results on other datasets are detailed in Sec. 7 of the Supplemental Material\cite{SM}. As a result, the edge-enhanced DNN architecture based on a single metasurface effectively improves classification performance while maintaining a simple single-layer structure.

\section{Conclusions}

In conclusion, we have demonstrated a single-layer nonlocal Huygens' metasurface that achieves all-optical edge detection preprocessing and DNN-based classification via time-sequential spin-multiplexing. Specifically, the co-polarized component in the output light exploits the dispersion properties of the nonlocal metasurface to perform edge detection. Subsequently, the cross-polarized component is modulated by the metasurface's geometric phase profile, enabling image classification within the DNN. By coupling q-BIC and MDR resonances to satisfy the generalized Kerker condition in crescent-shaped nanopillars, we achieve a high polarization conversion efficiency of approximately 55\%. The edge-enhanced DNN effectively elevates the classification accuracy of the single-layer network on the MNIST dataset from 64.2\% to 80.7\%. This integrated metasurface-based all-optical computing architecture provides a novel solution for the development of compact intelligent optical systems in applications such as autonomous driving perception, biomedical imaging, and industrial defect inspection.

\begin{acknowledgments}
	
This work was supported by the National Natural Science Foundation of China (Grants No. 12364045, No. 12304420, and No. 12264028), the Natural Science Foundation of Jiangxi Province (Grants No. 20262BAC240253 and No. 20253BAC260002), the Young Elite Scientists Sponsorship Program by JXAST (Grants No. 2023QT11 and No. 2025QT04), and the Innovation Fund for Graduate Students of Jiangxi Province (Grant No. YC2026-S119).

\end{acknowledgments}

\section*{Data availability}

The data that support the findings of this article are not publicly available. The data are available from the authors upon reasonable request.


%

\end{document}